\documentclass[modern]{aastex62}

\usepackage{natbib}
\usepackage{placeins}
\usepackage{rotfloat}
\usepackage{gensymb}
\usepackage{cancel}
\usepackage{amsmath}

\makeatletter
\def \blfootnote{\xdef\@thefnmark{}\@footnotetext}
\makeatother

\shorttitle{sofia fir polarimetric observations of 30~Dor}
\shortauthors{SOFIA Community Science I: 30 Dor}

\begin{document}

\title{SOFIA Community Science I: HAWC+ Polarimetry of 30 Doradus}

\correspondingauthor{SOFIA Helpdesk}
\email{sofia\_helpdesk@sofia.usra.edu}

\author[0000-0002-1913-2682]{M.~S.~Gordon}
\affiliation{SOFIA Science Center, NASA Ames Research Center, Moffett Field, CA 94035, USA}

\author[0000-0001-5357-6538]{E.~Lopez-Rodriguez}
\affiliation{SOFIA Science Center, NASA Ames Research Center, Moffett Field, CA 94035, USA}

\author{B.-G.~Andersson}
\affiliation{SOFIA Science Center, NASA Ames Research Center, Moffett Field, CA 94035, USA}

\author{M.~Clarke}
\affiliation{SOFIA Science Center, NASA Ames Research Center, Moffett Field, CA 94035, USA}

\author[0000-0002-0859-0805]{S.~Coud\'e}
\affiliation{SOFIA Science Center, NASA Ames Research Center, Moffett Field, CA 94035, USA}

\author[0000-0002-9820-1032]{A.~Moullet}
\affiliation{SOFIA Science Center, NASA Ames Research Center, Moffett Field, CA 94035, USA}

\author{S.~N.~Richards}
\affiliation{SOFIA Science Center, NASA Ames Research Center, Moffett Field, CA 94035, USA}

\author[0000-0002-7261-6504]{R.~Y.~Shuping}
\affiliation{SOFIA Science Center, NASA Ames Research Center, Moffett Field, CA 94035, USA}
\affiliation{Space Science Institute, 4750 Walnut Street, Suite 205, Boulder, CO 80301, USA}

\author{W.~Vacca}
\affiliation{SOFIA Science Center, NASA Ames Research Center, Moffett Field, CA 94035, USA}

\author{H.~Yorke}
\affiliation{SOFIA Science Center, NASA Ames Research Center, Moffett Field, CA 94035, USA}



\begin{abstract}

The Stratospheric Observatory for Infrared Astronomy (SOFIA) is a Boeing 747SP aircraft modified to accommodate a 2.7 meter gyro-stabilized telescope, which is mainly focused to studying the Universe at infrared wavelengths. As part of the Strategic Director's Discretionary Time (S-DDT) program, SOFIA performs observations of relevant science cases and immediately offers science-ready data products to the astronomical community. We present the first data release of the S-DDT program on far-infrared imaging polarimetric observations of \object[30 Doradus]{30~Doradus} using the High-resolution Airborne Wideband Camera-Plus (HAWC+) at 53, 89, 154, and 214~\micron. We present the status and quality of the observations, an overview of the SOFIA data products, and examples of working with HAWC+ polarimetric data that will enhance the scientific analysis of this, and future, data sets. These observations illustrate the potential influence of magnetic fields and turbulence in a star-forming region within the Tarantula Nebula.

\end{abstract}




\section{Preamble} \label{sec:pre}

The Stratospheric Observatory for Infrared Astronomy (SOFIA) is dedicated to studying the Universe at infrared wavelengths. The Strategic Director's Discretionary Time (S-DDT) for SOFIA is aimed at providing the astronomical community with data sets of high scientific interest over a broad range of potential research topics without any proprietary period.  These observations allow the general user community access to high-level data products that are meant not only for general understanding of SOFIA data and its packaging but also for inclusion in published scientific work.  The S-DDT targets are selected the basis of non-interference with existing programs.

The SDDT program 76\_0001, ``Community Science: HAWC+ Polarimetry of 30~Dor," was designed and scheduled to provide the community with SOFIA polarimetry data of an important and relatively bright source.  The observing strategy also provided for increased scheduling efficiency for the OC6I (HAWC+) flights in July~2018.  The west-bound observing legs for 30~Doradus allowed a larger fraction of the highest ranked Cycle 6 targets, predominantly in the inner Galaxy, to be scheduled and flown.

To enhance the scientific exploitation of these data products, we present here an overview of the observations, visualizations of the data, and a preliminary analysis of their quality.  Finally, this document and the accompanying \textsc{jupyter} notebook\footnote{HAWC+ \textsc{jupyter} notebook found at: \\ \url{https://nbviewer.jupyter.org/github/SOFIAObservatory/Recipes/blob/master/HAWC_30Dor.ipynb}} present basic tools for analysis of this dataset. These tools should be useful for the analysis of future SOFIA data releases as well.


\section{Observations \& Data Reduction} \label{sec:obs}

30~Doradus, the Tarantula Nebula ($\alpha_{2000}$:$\,05^\mathrm{h}38^\mathrm{m}38^\mathrm{s}$, $\delta_{2000}$:$\,-69^\circ05'42''$), was observed during the SOFIA New Zealand deployment as part of the S-DDT (PI: Yorke,~H, ID: 76\_0001) using the High-resolution Airborne Wideband Camera-Plus \citep[HAWC+; ][]{harper2018} on the SOFIA telescope. We performed imaging polarimetric observations in the wavelength range of 50--250~$\micron$ with HAWC+.

 Radiation, prior to entering the HAWC+ cryostat, passes through a set of warm, $\sim$220~K, fore-optics, i.e.\ radiation is reflected from a folding mirror to a field mirror, which allows for imaging the SOFIA pupil inside the HAWC+ cryostat. After the fore-optics, radiation enters the cryostat through a 7.6~cm diameter high-density polyethylene (HDPE) window, then passing through a cold pupil on a rotating carousel with near-infrared filters that define each bandpass, and finally lenses designed to optimize the plate scale. Table~\ref{tab:obs} summarizes the central wavelength and bandwidth of each band. The pupil carousel and the filter wheel are at a temperature of $\sim$4~K. The carousel contains eight aperture positions, four of which contain half wave plates (HWPs) for the four HAWC+ bands, an open aperture with a diameter matching the SOFIA pupil, and three aperture options for instrumental calibration. After the pupil carousel, the radiation passes through a wire grid that reflects one component of linear polarization and transmits the orthogonal component to the detector arrays. HAWC+ polarimetric observations simultaneously measure two orthogonal components of linear polarization arranged in a pair of arrays with $32\times40$ pixels. For further details on the HAWC+ instrument, we refer to the \citet{harper2018} instrument design and hardware paper.

\begin{deluxetable}{rccccccc}[htp]
  \caption{HAWC+ 30 Doradus Observation Log}
  \label{tab:obs}
  \tablewidth{0pt}
  \tablehead{\colhead{$\lambda_c$ (label)} & \colhead{Bandwidth} & \colhead{Pixel scale} & \colhead{Beam size} & \colhead{Date} & \colhead{Chop-throw} & \colhead{Chop-angle} & \colhead{Obs. time}  \\
    \colhead{($\micron$)} & \colhead{($\micron$)} & \colhead{($\arcsec$)} & \colhead{($\arcsec$)} & \colhead{yyyy/mm/dd} & \colhead{($\arcsec$)} & \colhead{($\arcsec$)} & \colhead{(s)}}
  \startdata
  53 (A) & 8.7 & 2.55 & 4.85 & 2018/07/12\phn & 480 &  120 & 904 \\
  89 (C) & 17 & 4.02 & 7.80 & 2018/07/07\tablenotemark{*} & 480 & 120 & 3700 \\
  154 (D) & 34 & 6.90 & 13.6 & 2018/07/05\phn & 480 & 120 & 2836 \\
  214 (E) & 44 & 9.37 & 18.2 & 2018/07/12\phn & 480 & 120 & 2222 \\
  \enddata
  \tablenotetext{*}{\scriptsize \textsc{Note} --- Observations in HAWC+ 89 \micron\ were taken between 2018/07/07 and 2018/07/11.  Level 4 data on SOFIA \textsc{dcs} represent the combined observation time of 3700~s between the two nights.}
\end{deluxetable}

Polarimetric observations were performed in the chop-nod observing mode together with a four-dither pattern with an offset of three pixels (8\arcsec, 12\arcsec, 21\arcsec, 28\arcsec\ for the 53~\micron, 89~\micron, 154~\micron, 214~\micron\ filters). Four half-wave plate position angles ($5^\circ$, $50^\circ$, $27.5^\circ$, and $72.5^\circ$) were observed in each dither position at a chop frequency of 10.2~Hz and nod times of 40~s. Table~\ref{tab:obs} summarizes the observation details and filter-specific information for the observations. The chop-throw and chop-angle configuration were chosen to avoid any significant flux contribution from the diffuse emission from the background.

The data were reduced using the \textsc{hawc\_drp pipeline} v1.3.0. A full description of the HAWC+ Pipeline supported by SOFIA can be found in the Data Handbook and the Pipeline Users' Manual available at the Data Resources SOFIA website.\footnote{HAWC+ Data Products handbook: \url{https://www.sofia.usra.edu/sites/default/files/Instruments/HAWC_PLUS/Documents/HAWC_GO_Handbook_RevB.pdf}} In brief, the raw data were chop-nod subtracted, flat fielded, corrected for telluric absorption, and flux calibrated. The frames for the individual dither positions were combined into a single mosaic.  Finally, the Stokes $IQU$ parameters and their uncertainties were estimated. In Section \ref{sec:data}, we outline the output HAWC+ data cube and present image visualization techniques.


\section{Data Products \& Handling HAWC+ Data}\label{sec:data}

\begin{figure}[h!]
  \includegraphics[scale = 0.33]{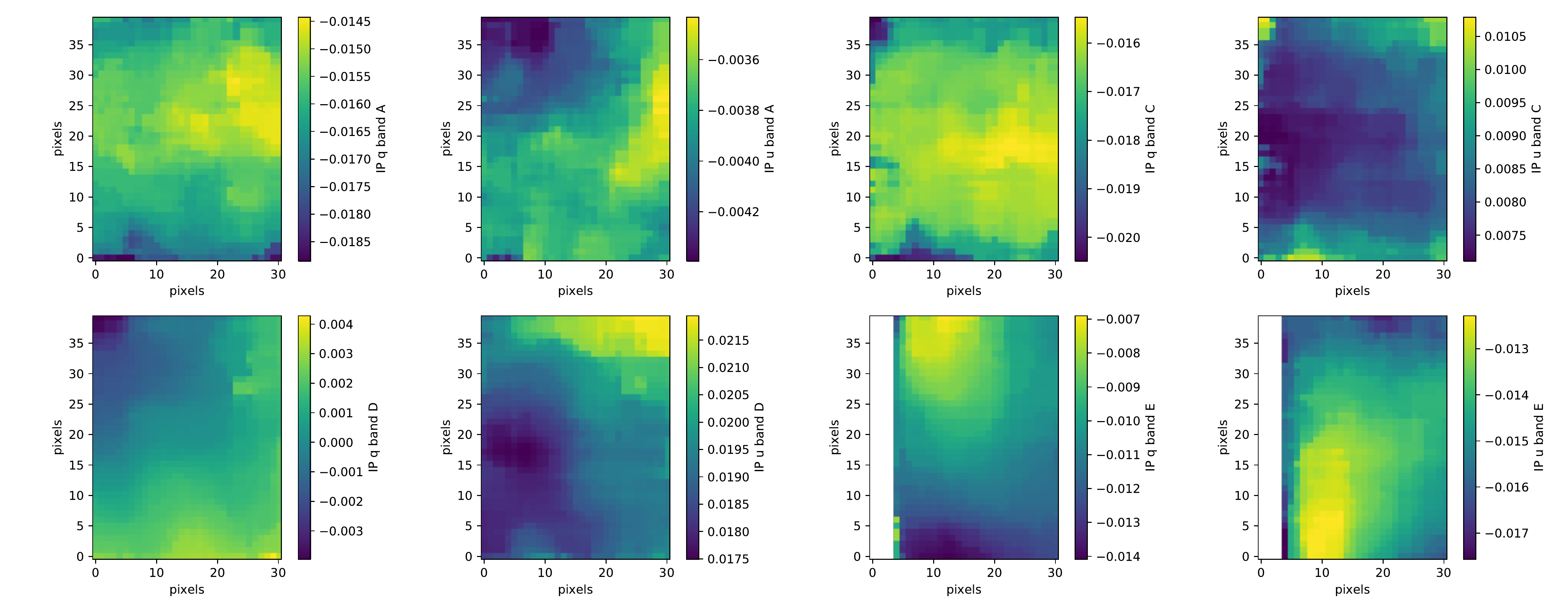}
  \caption{Instrumental polarization of the normalized Stokes $QU$ at 53~\micron, 89~\micron, 154~\micron, and 214~\micron. Instrumental polarization ranges from $1.2-1.8$\%.
    \label{fig:ip}}
\end{figure}

The data pipeline group at SOFIA Science Center delivers HAWC+ science-ready products at several reduction levels. Data delivered at `Level 3' is flux calibrated, corrected for instrumental polarization, polarization bias, polarization efficiency, and zero-angle. Flux calibration is performed during every observing campaign with HAWC+ using asteroids and/or planets. We estimate a flux calibration accuracy of better than $10\%$. Instrumental polarization is estimated to be in the range of $1.2-1.8\%$ with a reproducibility of $\le 0.3\%$ between different epochs. Figure~\ref{fig:ip} shows the instrumental polarization maps of the normalized Stokes $QU$ at each HAWC+ band applied during the data reduction. Note that observations at 214~$\micron$ suffer from internal vignetting across five columns on the left of the array. These columns are masked during the data reduction, and the instrumental polarization maps at 214~$\micron$ show the five columns as blank. The instrumental polarization in all bands primarily originates from the SOFIA tertiary mirror, with a smaller contribution from HAWC+ itself. The polarization efficiency is estimated at $\ge 84$\% for all bands. The zero-angle calibration is performed using internal polarization calibrators located in the fore-optics which ensure an absolute polarization angle uncertainty of better than $3^{\circ}$. For more details on these calibrations procedures see \citet{harper2018}. 

HAWC+ `Level~4' data corresponds to fully calibrated data combined from different observing nights. The Level 4 data products are \textsc{fits} files each containing 19 extensions.  Table~\ref{tab:ext} summarizes each Header Data Unit (\textsc{hdu}) in the \textsc{fits} file structure. Every dataset contains the Stokes $IQU$, fractional polarization ($p$), position angle (PA) of polarization ($\theta$), and polarized flux ($I_p$), as well as their associated uncertainties. Final data products are delivered with a pixel scale equal to the Nyquist sampling at each band, i.e.\ 1\farcs27, 2\farcs01, 3\farcs45, and 4\farcs68 at 53~\micron, 89~\micron, 154~\micron, and 214~\micron, respectively. During the data reduction, a Gaussian kernel with a full width at half-maximum (FWHM) equal to the detector pixel scale was used to smooth, resample, and merge the observations at each dither position.

\begin{deluxetable}{c|l|c|c|l}[htp]
  \caption{HAWC+ Level 4 \textsc{fits} Extensions}
  \label{tab:ext}
  \tablewidth{0pt}
  \tablehead{\colhead{Ext \#} & \colhead{Ext Name} & \colhead{Type} & \colhead{Units} & \colhead{Description}}
  \startdata
  \phn0 & \textsc{stokes i} & img & Jy/pix & Stokes $I$ (total intensity)\\
  \phn1 & \textsc{error i} & img & Jy/pix & Error in $I$\\
  \phn2 & \textsc{stokes q} & img & Jy/pix & Stokes $Q$ \\
  \phn3 & \textsc{error q} & img & Jy/pix & Error in $Q$\\
  \phn4 & \textsc{stokes u} & img & Jy/pix & Stokes $U$ \\
  \phn5 & \textsc{error u} & img & Jy/pix & Error in $U$\\
  \phn6 & \textsc{image mask} & img & \nodata & Weighted \# of input pixels combined into output pixels\\
  \phn7 & \textsc{percent pol} & img & \% &  Polarization percent $p=100\sqrt{\left(Q/I\right)^2+\left(U/I\right)^2}$\\
  \phn8 & \textsc{debiased percent pol} & img & \% & Debiased polarization percent $p^\prime=\sqrt{p^2-\sigma_p^2}$\\
  \phn9 & \textsc{error percent pol} & img & \% & Error in $p^\prime$\\
  10 & \textsc{pol angle} & img & deg & Polarization angle ($\theta$) in sky coordinates\\
  11 & \textsc{rotated pol angle} & img & deg & Polarization angle ($\theta_{90}$) rotated by $90^{\circ}$\\
  12 & \textsc{error pol angle} & img & deg & Error in $\theta$ \\
  13 & \textsc{pol flux} & img & Jy/pix & Polarized intensity $I_p=I\times p/100$\\
  14 & \textsc{error pol flux} & img & Jy/pix & Error in $I_p$\\
  15 & \textsc{debiased pol flux} & img & Jy/pix & Debiased polarized intensity $I_{p^\prime}=I\times p^\prime/100$\\
  16 & \textsc{merged data} & tab & \nodata & Detector info from all merged images in cube\\
  17 & \textsc{pol data}\tablenotemark{*} & tab & \nodata & Polarization data for each pixel\\
  18 & \textsc{final pol data}\tablenotemark{$\dagger$} & tab & \nodata & Subset of \textsc{pol data} with quality cuts\\
  \enddata
  \tablenotetext{*}{\scriptsize \textsc{pol data} is a table representation of the $\theta$, $p$, and $p^\prime$ maps in extensions 7--12 for every pixel in both \textsc{x},\textsc{y} and \textsc{ra},\textsc{dec} coordinates. See Table~\ref{tab:poldata} for a description of the table columns.}
  \tablenotetext{\dagger}{\scriptsize \textsc{final pol data} is a quality cut of \textsc{pol data} for all pixels with $I/\sigma_I\geq200$, $p/\sigma_p\geq3$, and $p<50\%$.}
\end{deluxetable}

\begin{figure}[h!]
  \includegraphics[scale=0.9]{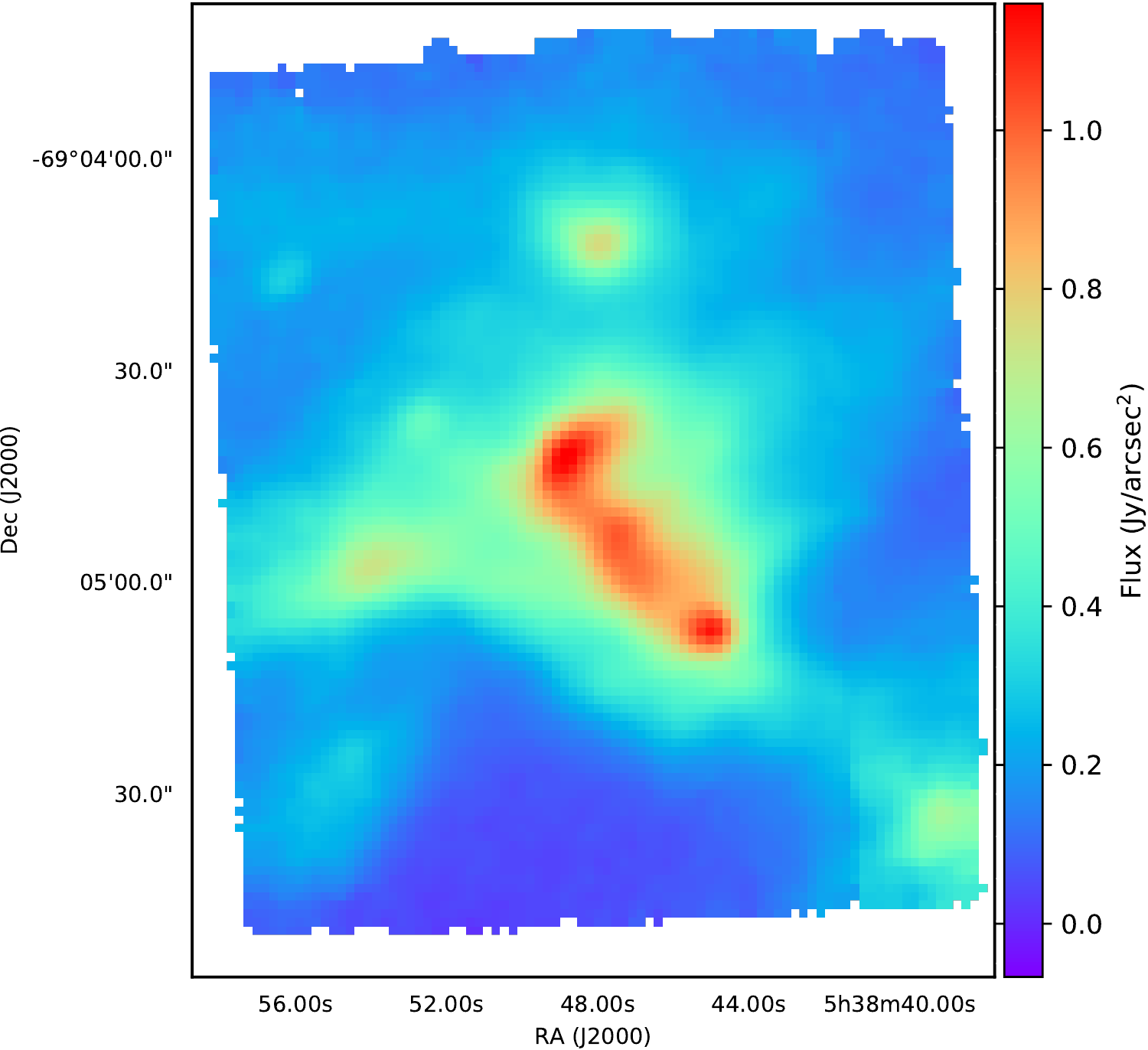}
  \caption{Total intensity (Stokes $I$) of 30~Dor at 53~\micron. Color scale shows the surface brightness in units of Jy/arcsec$^2$.
    \label{fig:stokes_i}}
\end{figure}

\begin{figure}[h!]
  \includegraphics[scale=0.52]{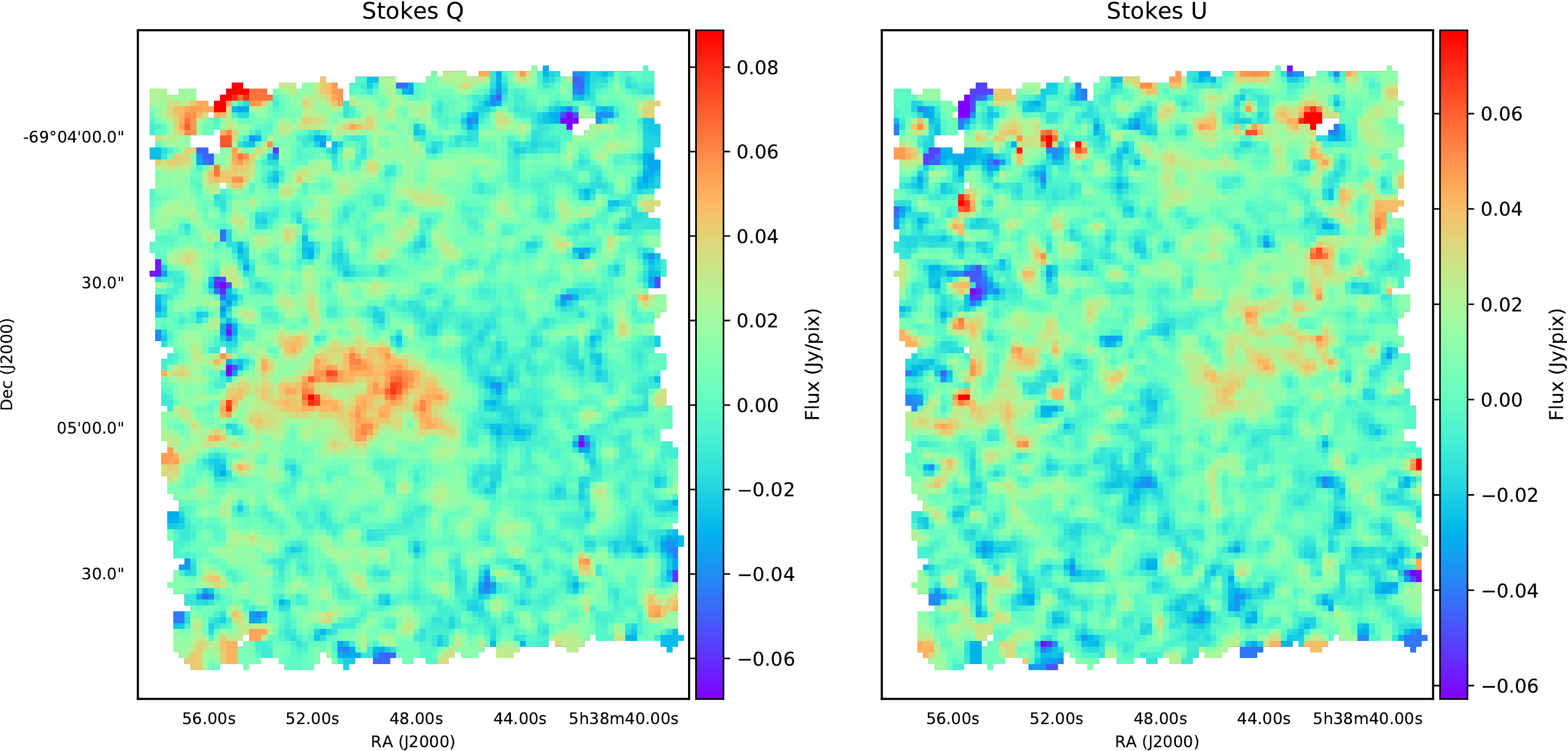}
  \caption{Stokes $Q$ and $U$ of 30 Dor at 53~\micron. Color scales show the flux density in units of Jy/pixel.
    \label{fig:stokes_q_u}}
\end{figure}

Here we show an example of these products for the observations of 30~Dor at 53~\micron. 30~Dor data can be downloaded from the SOFIA Data Cycle system (\textsc{dcs})\footnote{30~Doradus data can be found at: \\ \url{https://www.sofia.usra.edu/multimedia/science-results-archive/sofia-reveals-never-seen-magnetic-field-details}}. The following data analysis and figures are performed using \textsc{python} and can be found on the SOFIA website as a \textsc{jupyter} notebook.\footnote{Also available on GitHub at: \\
\url{https://github.com/SOFIAObservatory/Recipes}} Figure~\ref{fig:stokes_i} shows the total intensity image Stokes $I$ of 30~Dor at 53~\micron. The individual Stokes $Q$ and $U$ maps can be found in Figure~\ref{fig:stokes_q_u}.

Data products at Level~4 provide extensions with the polarization fraction ($p$), angle ($\theta$), and their associated errors ($\sigma$). Percent polarization and error are calculated as:

\begin{equation}\label{eq:p}
  p = 100\sqrt{\left(\frac{Q}{I}\right)^2+\left(\frac{U}{I}\right)^2}
\end{equation}

\begin{equation}\label{eq:sP}
\begin{split}
  \sigma_p  &=  \frac{100}{I} \biggl\{ \frac{1}{(Q^2+U^2)}\left[(Q\,\sigma_Q)^2+(U\,\sigma_U)^2+2QU\,\sigma_{QU}\right]  \\  
&+ \left[\left(\frac{Q}{I}\right)^2 + \left(\frac{U}{I}\right)^2\right]\sigma_I^2- 2\frac{Q}{I}\sigma_{QI}-2\frac{U}{I}\sigma_{UI} \biggr\}^{\!1/2}
\end{split}
\end{equation}
\noindent \\
Note that $p$ represents the percent polarization as opposed to fractional polarization, and that the uncertainty in the degree of polarization incorporates all covariance terms. Maps of these data are found in extensions 7 (\textsc{percent pol}) and 9 (\textsc{error percent pol}). The debiased polarization percentage ($p^\prime$), found in extension 8 (\textsc{debiased percent pol}), is calculated as:

\begin{equation}\label{Pb}
  p^\prime=\sqrt{p^2-\sigma_p^2},
\end{equation}
following the approach by \citet{wardle1974}, where the level of polarization bias depends on the signal-to-noise ratio (SNR) of the measurements. The lower the SNR, the larger the uncertainty in the degree of polarization and thus the lower the accuracy. We strongly recommend using the debiased polarization (extension 8, \textsc{debiased percent pol}) for any polarimetric analysis. For more information see \cite{serkowski1958,wardle1974,simmons1985}.

Polarized intensity, $I_p$, can then be calculated as $I_p = \frac{I\times p}{100}$, which is included in extension 13 (\textsc{pol flux}). We similarly define the debiased polarized intensity as $I_{p^\prime} = \frac{I\times p^\prime}{100}$, included in extension 15 (\textsc{debiased pol flux}).

Extensions 7-12 are also provided as a table and can be found in extension 17 (\textsc{pol data}), which lists every pixel with \textsc{(x,y)} and \textsc{(ra,dec)} coordinates, as well as the polarization data. Table \ref{tab:poldata} summarizes the entries of extension 17 (\textsc{pol data}).

\begin{figure}[h!]
    \includegraphics[scale=0.55]{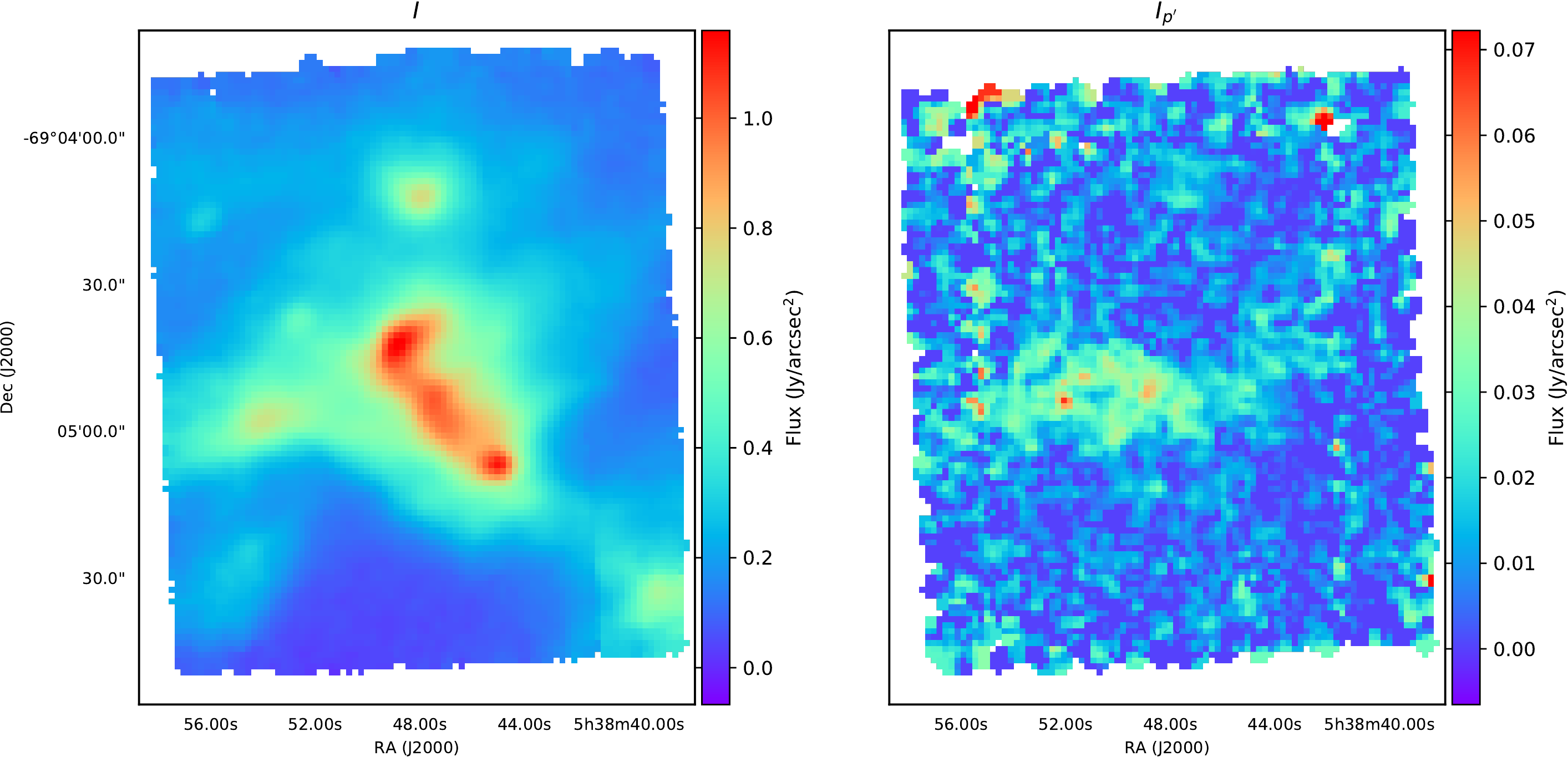}
  \caption{Stokes $I$ and debiased polarized flux $I_p^\prime$ of 30~Dor at 53~\micron. Color scales in units of Jy/arcsec$^{2}$.
    \label{fig:stokes_ip}}
\end{figure}

\begin{deluxetable}{l|c|l}[htp]
  \caption{\textsc{pol data} Table Description}
  \label{tab:poldata}
  \tablewidth{0pt}
  \tablehead{\colhead{Column Name} & \colhead{Units} & \colhead{Description}}
  \startdata
  Pixel X & pix & re-sampled pixels coordinates\\
  Pixel Y & pix & re-sampled pixel coordinates \\
  Right Ascension & deg & WCS coordinates \\
  Declination & deg & WCS coordinates \\
  Percent Pol & \% & Polarization percent $p$, extension 7 (\textsc{percent pol}) \\
  Debiased Percent Pol & \% & Debiased polarization percent $p^\prime$, extension 8 (\textsc{debiased percent pol}) \\
  Err. Percent Pol & \% & Error in debiased polarization $p^\prime$, extension 9 (\textsc{error percent pol}) \\
  Theta & deg & Polarization angle ($\theta$) in sky coordinates, extension 10 (\textsc{pol angle}) \\
  Rotated Theta & deg & Polarization angle rotated by 90$^{\circ}$ ($\theta_{90}$) , extension 11 (\textsc{rotated pol angle}) \\
  Err. Theta & deg & Error in $\theta$, extension 12 (\textsc{error pol angle}) \\
  \enddata
\end{deluxetable}


\section{Polarization maps}
This section shows how to produce polarization maps using Level~4 data products generated by the HAWC+ data reduction pipeline. From the Stokes $Q$ and $U$ maps, the polarization angle $\theta$ is calculated in the standard manner:
\begin{equation}\label{eq:q}
  \theta = \frac{90}{\pi}\,\mathrm{tan}^{-1}\left(\frac{U}{Q}\right)
\end{equation}
\noindent
with associated error:
\begin{equation}
  \sigma_\theta = \frac{90}{\pi\left(Q^2+U^2\right)}\sqrt{\left(Q\sigma_U\right)^2+\left(U\sigma_Q\right)^2-2QU\sigma_{QU}}
\end{equation}

As mentioned in Section \ref{sec:data}, the PA of polarization map is stored in extension 10 (\textsc{pol angle}) with its error in extension 12 (\textsc{error pol angle}). Note that, as part of the HAWC+ reduction pipeline, $\theta$ is already corrected for the zero-angle of polarization and is provided in sky coordinates, such that $\theta = 0^{\circ}$ corresponds to the North-South direction, $\theta = 90^{\circ}$ corresponds to the East-West direction, and positive values correspond to counterclockwise rotation.

We also provide the PA of polarization rotated by $90^{\circ}$, $\theta_{90}$, in extension 11 (\textsc{rotated pol angle}). This quantity should be used with caution. If the measured polarization is dominated by magnetically-aligned dust grains, then the PA of polarization, $\theta$, can be rotated by $90^{\circ}$ to visualize the magnetic field morphology. For more details see \citet{hildebrand2000,andersson2015}.

We can now use the $p^\prime$ and $\theta_{90}$ maps to plot the polarization vectors. Before we plot the polarization map we make several quality cuts to ensure that we are plotting only physically-significant polarization vectors. First, we select polarization vectors with a signal-to-noise ratio (SNR) in the degree of polarization of $p'/\sigma_{p}>3$. For polarization vectors below this cut, polarization bias would dominate (see Section \ref{sec:data}). Second, we produce a cut on highly polarized vectors, i.e.\ $p' > 50\%$. Finally, we make a cut in the SNR of the total intensity ($SNR_{I} = I/\sigma_I$) which we show below is tied to the uncertainty in the degree of polarization for typical observations dominated by shot noise.

Starting with the propagated error on the fractional polarization (Equation~\ref{eq:sP}), we assume that the errors in Stokes $Q$ and $U$ are similar, $\sigma_Q \sim \sigma_U$, and we label them as $\sigma_{Q,U}$.  We additionally assume that the covariants (cross terms) of Stokes $IQU$ are negligible, i.e.\ $\sigma_{QU} \sim \sigma_{QI} \sim \sigma_{UI} \sim 0$. Therefore, Equation~\ref{eq:sP} can be written in the form:
\begin{equation}
\sigma_p  = \frac{1}{I}\sqrt{\sigma_{Q,U}^2+\sigma_I^2\left(\frac{Q^2+U^2}{I^2}\right)}=
\frac{1}{I}\sqrt{\sigma_{Q,U}^2+\sigma_I^2\,p^2}
\end{equation}

By design, the HAWC+ optics split the incident radiation into two orthogonal linear polarization states that are measured with two independent detector arrays. The total intensity, Stokes~$I$, is recovered by linearly adding both polarization states. If the data is taken at four equally-spaced HWP angles, and assuming 100\% efficiency of the instrument, then the uncertainty in the fractional polarization has the form
\begin{equation}\label{eq:sP1}
    \sigma_p = \frac{\sigma_I \sqrt{2}}{I}\left(1 + \frac{p^2}{2} \right)^{1/2}
\end{equation}
\begin{equation}\label{eq:sPSNR}
    \sim \frac{\sqrt{2}\sigma_I}{I} = \frac{\sqrt{2}}{\mathrm{SNR}_I}    
\end{equation}
In general, the fractional polarization, $p$, is relatively small; therefore, the last term of equation \ref{eq:sP1} can be ignored, leading to equation \ref{eq:sPSNR} where SNR$_{I}$ is the SNR in Stokes~$I$.

Thus, if we desire an uncertainty in the degree of polarization of $\sim0.5\%$, the required SNR in Stokes~$I$ is given by

\begin{equation}\label{eq:SNRI_sP}
    \left(\mathrm{S/N}\right)_I  \sim \sqrt{2}\left(\frac{1}{\sigma_p}\right) \sim \sqrt{2}\frac{1}{0.5\%}  \sim \frac{\sqrt{2}}{0.005} \sim 283
\end{equation}
\noindent
Therefore, to obtain an uncertainty in the degree of polarization of $\sigma_p\sim0.5\%$, we require a SNR in Stokes~$I$ of $\sim 283$. Note that this approach assumes a 100\% efficiency of polarization, 100\% time on-source, and no systematic errors. The estimated SNR in Stokes~$I$ can then be used as a quality cut on the polarization maps to both avoid noisy vectors and guarantee a specified uncertainty in $p$.

Level~4 products are delivered with quality cuts in the \textsc{hdu} labeled as \textsc{final pol data}. This extension contains a table similar to \textsc{pol data} with polarization vectors with the following quality cuts:
\begin{enumerate}
\item  $I/\sigma_I>200$
\item $p/\sigma_p>3$
\item  $p<50\%$
\end{enumerate}

\begin{figure}[h!]
  \includegraphics[scale = 0.47]{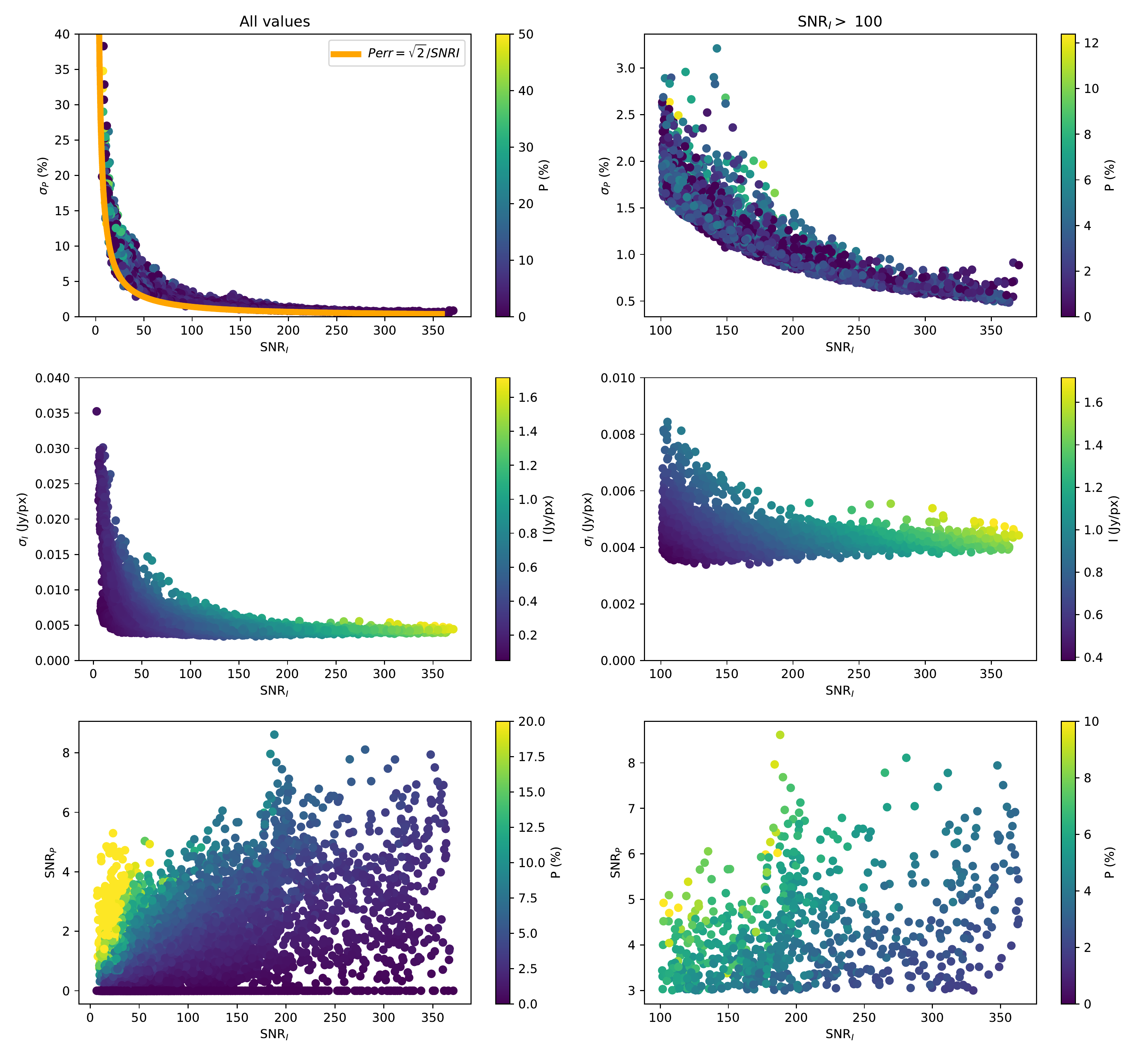}
  \caption{Summary of quality cuts performed for the polarization maps. These data represent the final polarization vectors used for the example map from Band A, 53~\micron. \textit{Top panel}: polarization uncertainty ($\sigma_p$) vs.\ SNR in Stokes $I$ (SNR$_I$) for all polarization vectors (left) and those  with a SNR$_{I} > 100$ (right). Equation~\ref{eq:sPSNR} is plotted as a reference (orange curve). \textit{Middle panel}: Stokes $I$ uncertainty ($\sigma_I$) vs.\ SNR$_{I}$ for all polarization vectors (left) and and those  with a SNR$_{I} > 100$ (right). \textit{Bottom panel}: SNR in the degree of polarization (SNR$_p$) vs.\ SNR in Stokes $I$ ($SNR_{I}$) for all polarization vectors (left) and those with SNR$_I > 100$ and SNR$_p > 3$ (right). This last plot shows all the polarization vectors plotted in Figure~\ref{fig:apolmap}. Figure~\ref{fig:PvsI} shows similar quality cuts for all four HAWC+ bands.
    \label{fig:sncuts}}
\end{figure}

\begin{figure}[h!]
  \includegraphics[scale = 0.85]{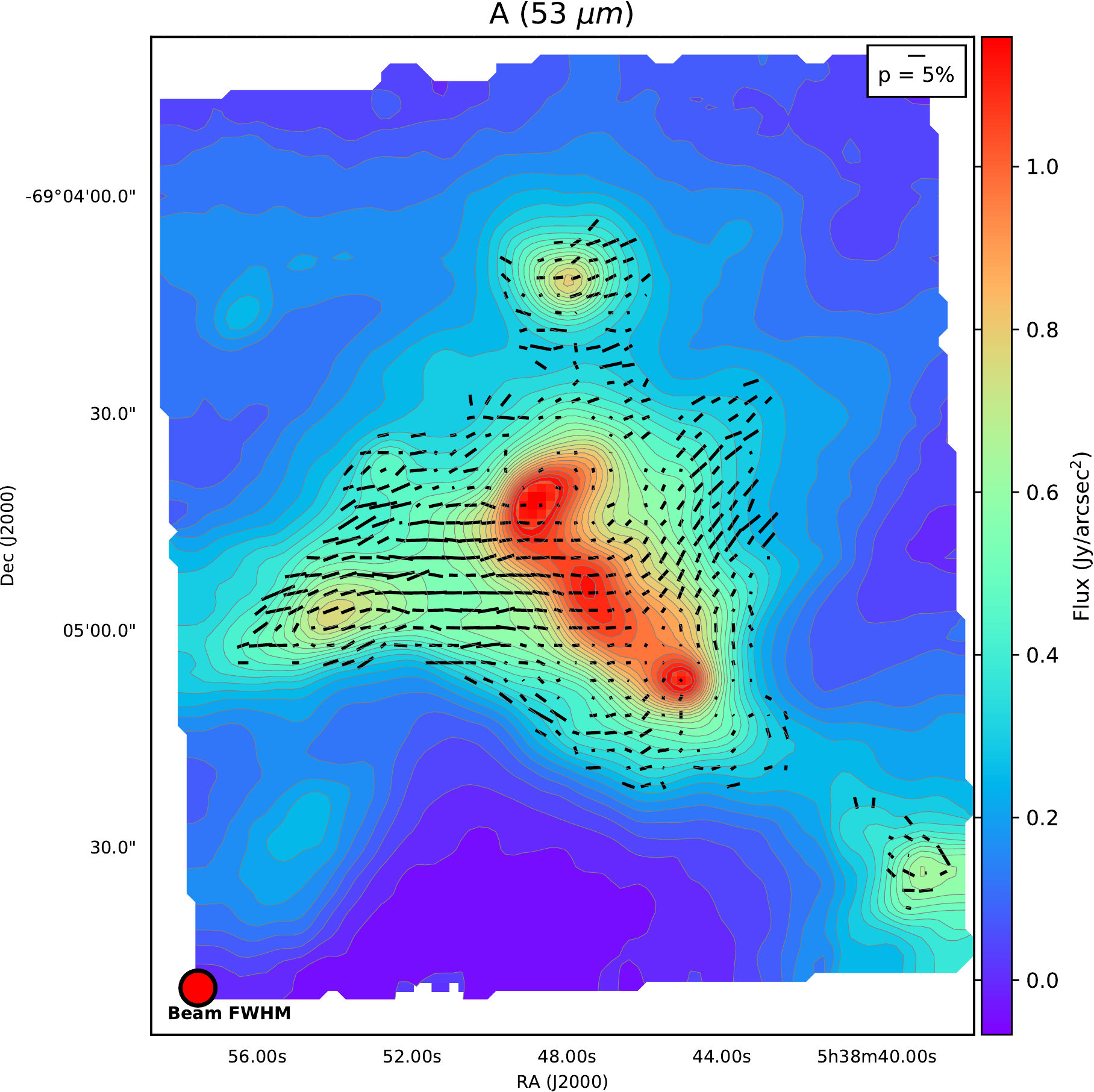}
  \caption{30 Dor polarization map at 53~\micron. Color scale shows the total intensity in units of Jy/arcsec$^2$. The overlaid polarization vectors (black lines) are proportional in length to the degree of polarization, and their orientation shows the PA of polarization rotated by $90^{\circ}$. The beam size and 5\% polarization are shown as references.
    \label{fig:apolmap}}
\end{figure}

\begin{figure}[h!]
  \includegraphics[scale = 0.85]{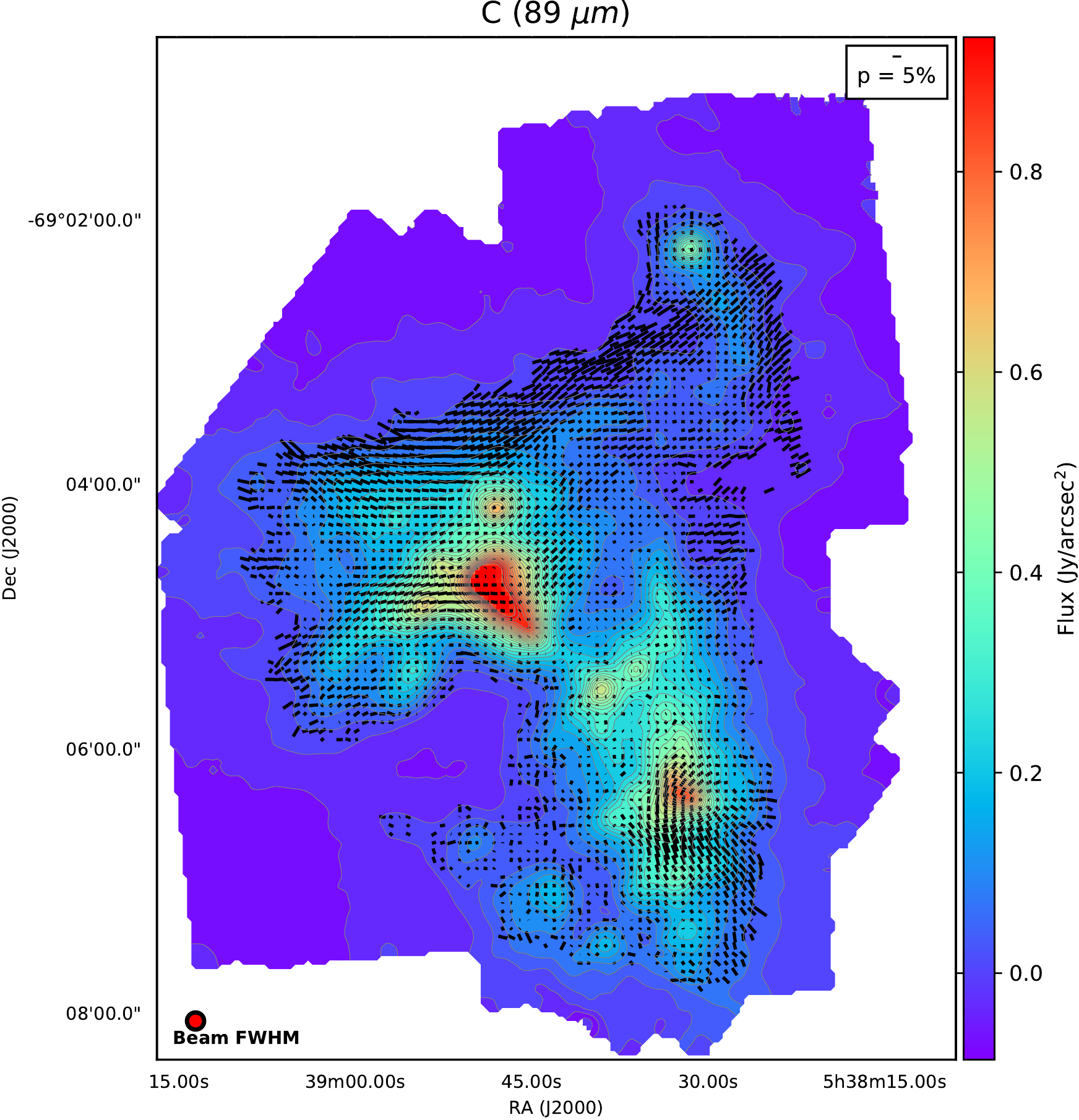}
  \caption{30 Dor polarization map at 89~\micron. Color scale shows the total intensity in units of Jy/arcsec$^2$. The overlaid polarization vectors (black lines) are proportional in length to the degree of polarization, and their orientation shows the PA of polarization rotated by $90^{\circ}$. The beam size and 5\% polarization are shown as references.
    \label{fig:cpolmap}}
\end{figure}

\begin{figure}[h!]
  \includegraphics[scale = 0.85]{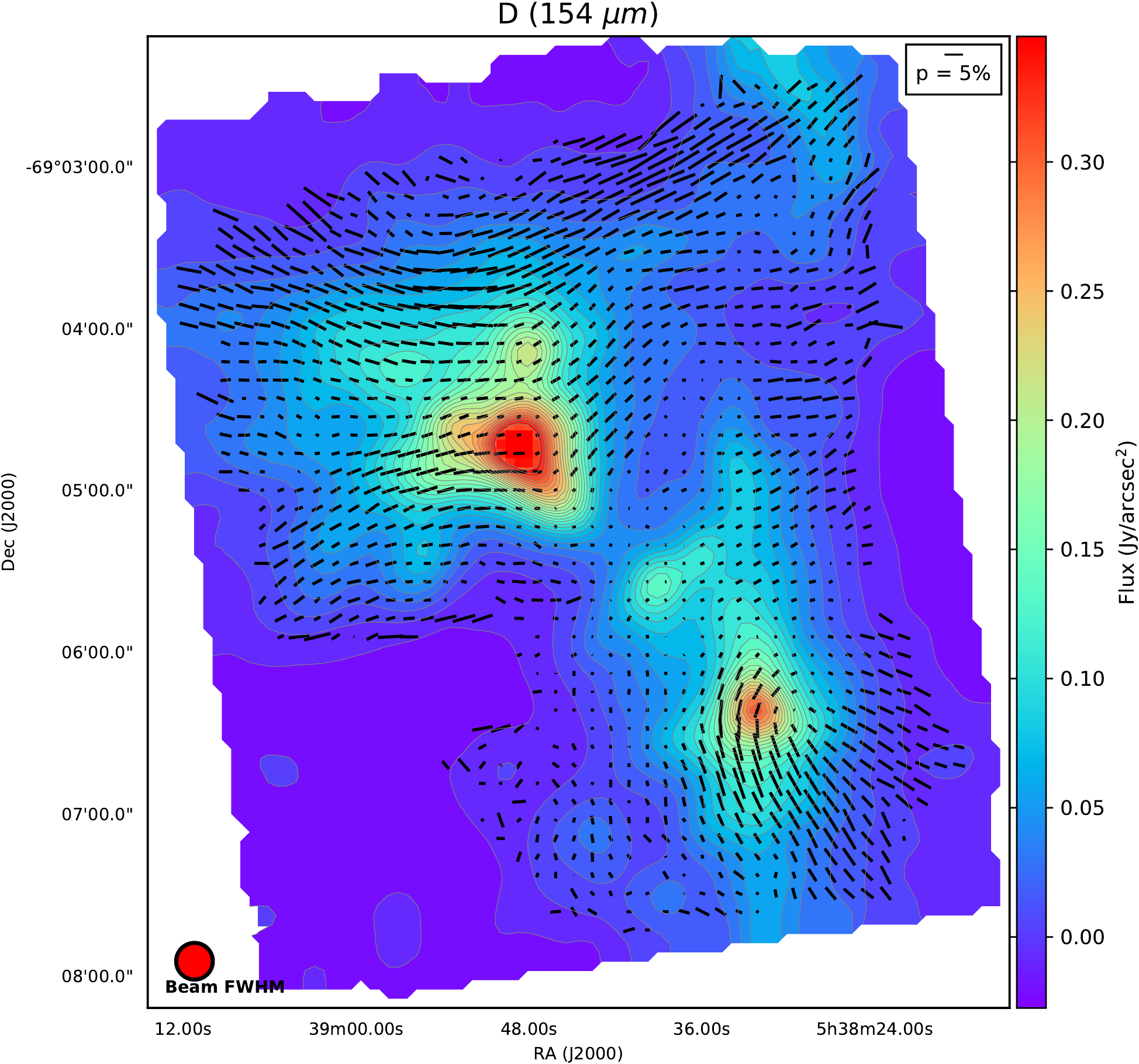}
  \caption{30 Dor polarization map at 154~\micron. Color scale shows the total intensity in units of Jy/arcsec$^2$. The overlaid polarization vectors (black lines) are proportional in length to the degree of polarization, and their orientation shows the PA of polarization rotated by $90^{\circ}$. The beam size and 5\% polarization are shown as references.
    \label{fig:dpolmap}}
\end{figure}

\begin{figure}[h!]
  \includegraphics[scale = 0.85]{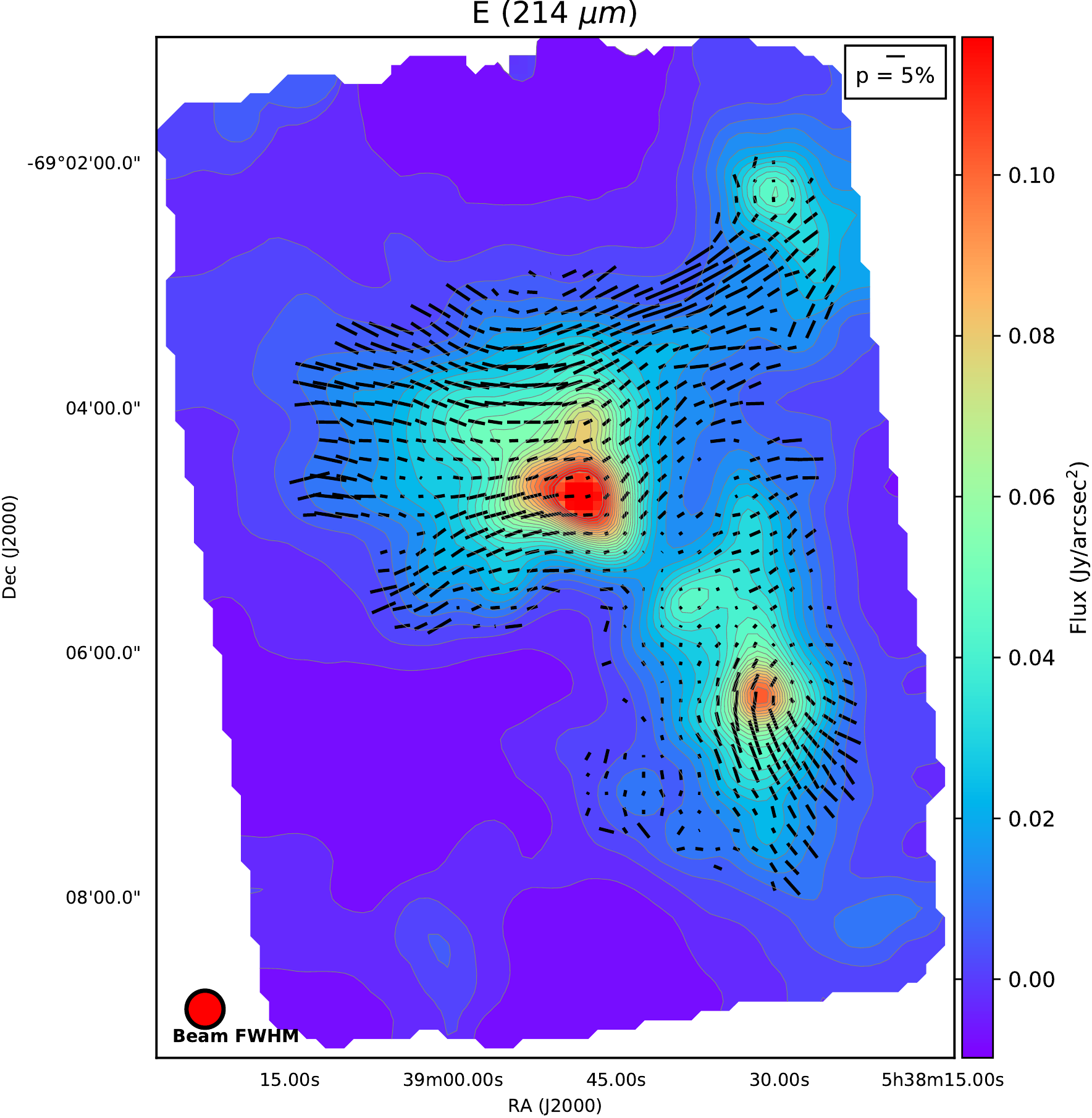}
  \caption{30 Dor polarization map at 214~\micron. Color scale shows the total intensity in units of Jy/arcsec$^2$. The overlaid polarization vectors (black lines) are proportional in length to the degree of polarization, and their orientation shows the PA of polarization rotated by $90^{\circ}$. The beam size and 5\% polarization are shown as references.
    \label{fig:epolmap}}
\end{figure}

These quality cuts are very conservative. We encourage investigators to decide on any quality cuts that satisfy their scientific needs. As examples, we produce polarization maps (Figures~\ref{fig:apolmap}--\ref{fig:epolmap}) of 30~Dor with a less restrictive cut of $\mathrm{SNR}_I > 100$, which corresponds to an uncertainty of the degree of polarization of 1.4\%. Figure~\ref{fig:sncuts} illustrates the quality cuts made for the polarization maps at 53 $\micron$.

In Figures~\ref{fig:apolmap}--\ref{fig:epolmap}, the overlaid polarization vectors are proportional in length to the degree of polarization, and their orientation shows the PA of polarization rotated by $90^{\circ}$ (i.e.\ roughly in the direction of the magnetic field). For all figures, only every other vector is shown for clarity. Note that the polarization maps have different sizes due to different fields-of-view for the four bands, as well as different dither sizes and multiple mapping positions joined together as mosaics.

\begin{figure}[h!]
  \includegraphics[scale = 0.85]{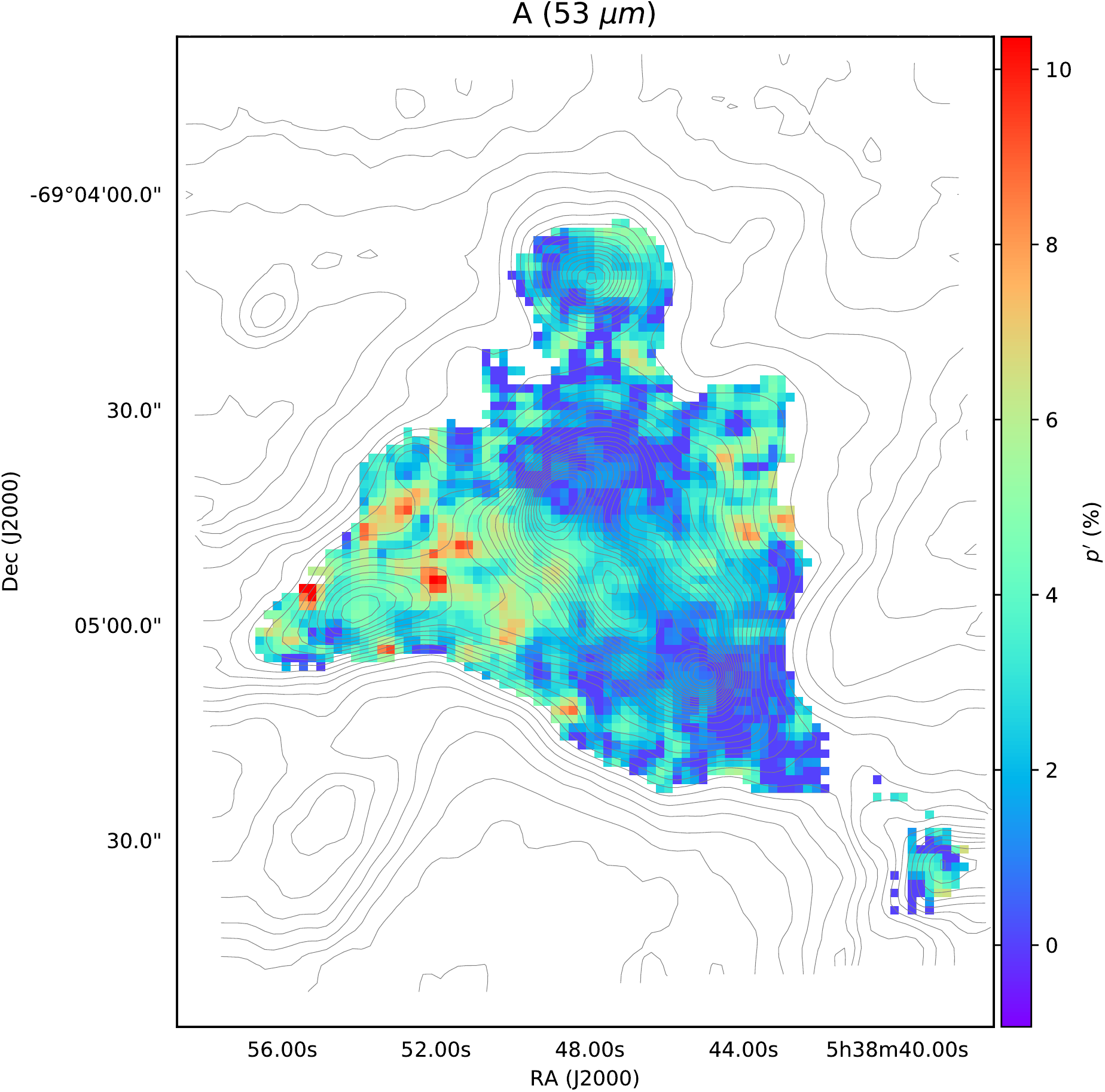}
  \caption{Percentage of polarization (color scale) at 53~\micron\ with overlaid total intensity contours.
    \label{fig:apmap}}
\end{figure}

\begin{figure}[h!]
  \includegraphics[scale = 0.85]{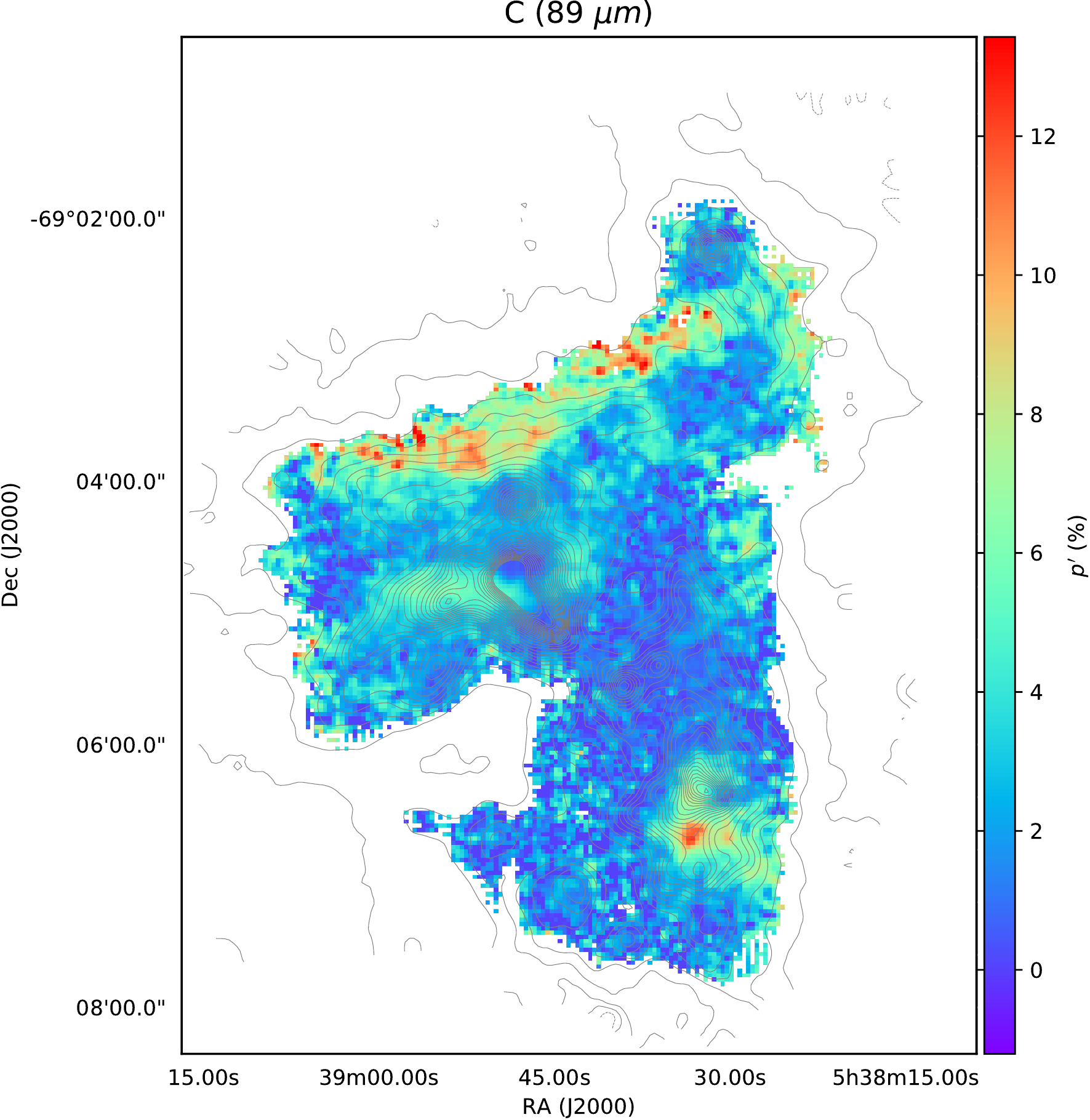}
  \caption{Percentage of polarization (color scale) at 89~\micron\ with overlaid total intensity contours.
    \label{fig:cpmap}}
\end{figure}

\begin{figure}[h!]
  \includegraphics[scale = 0.85]{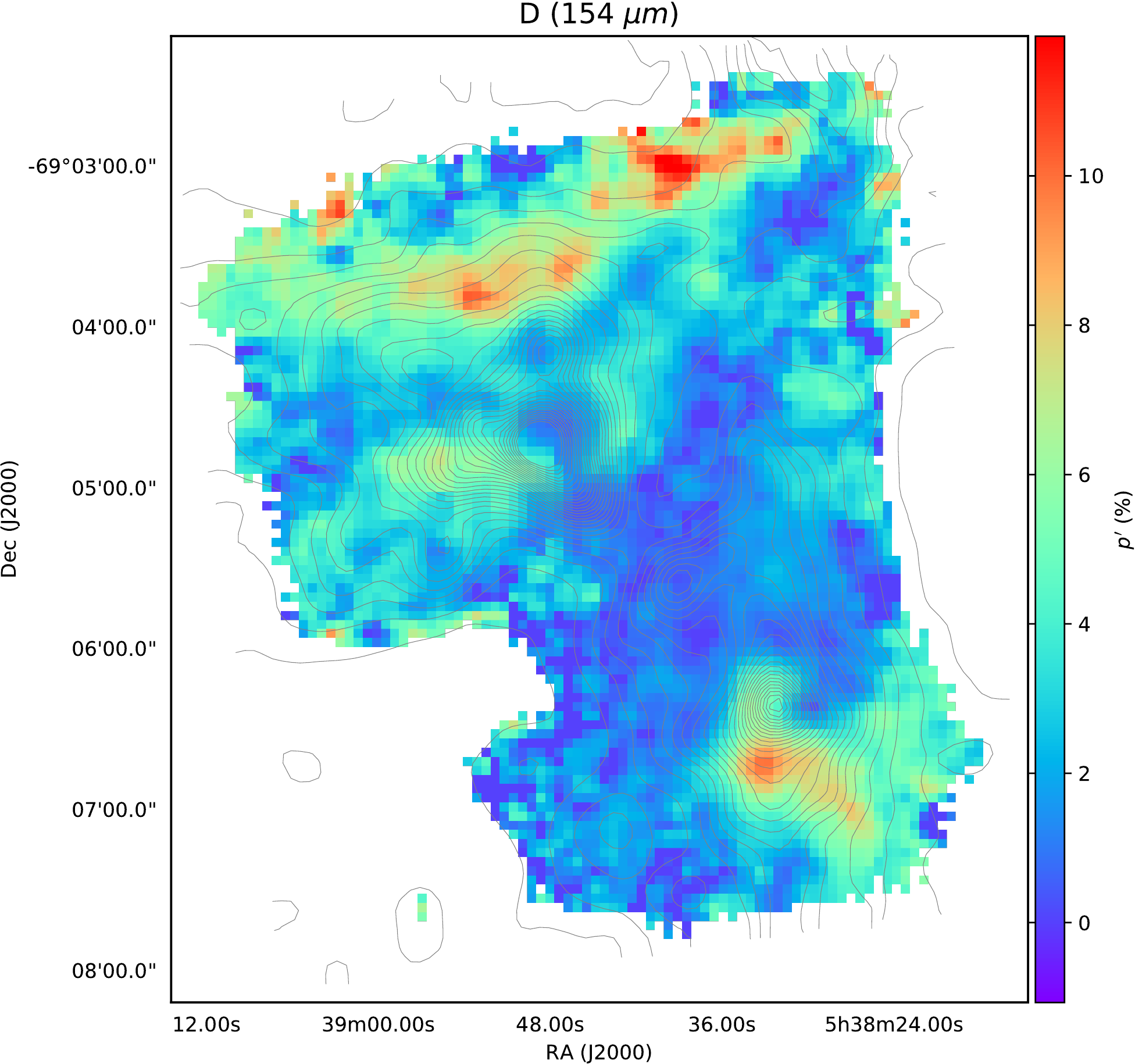}
  \caption{Percentage of polarization (color scale) at 154~\micron\ with overlaid total intensity contours.
    \label{fig:dpmap}}
\end{figure}

\begin{figure}[h!]
  \includegraphics[scale = 0.85]{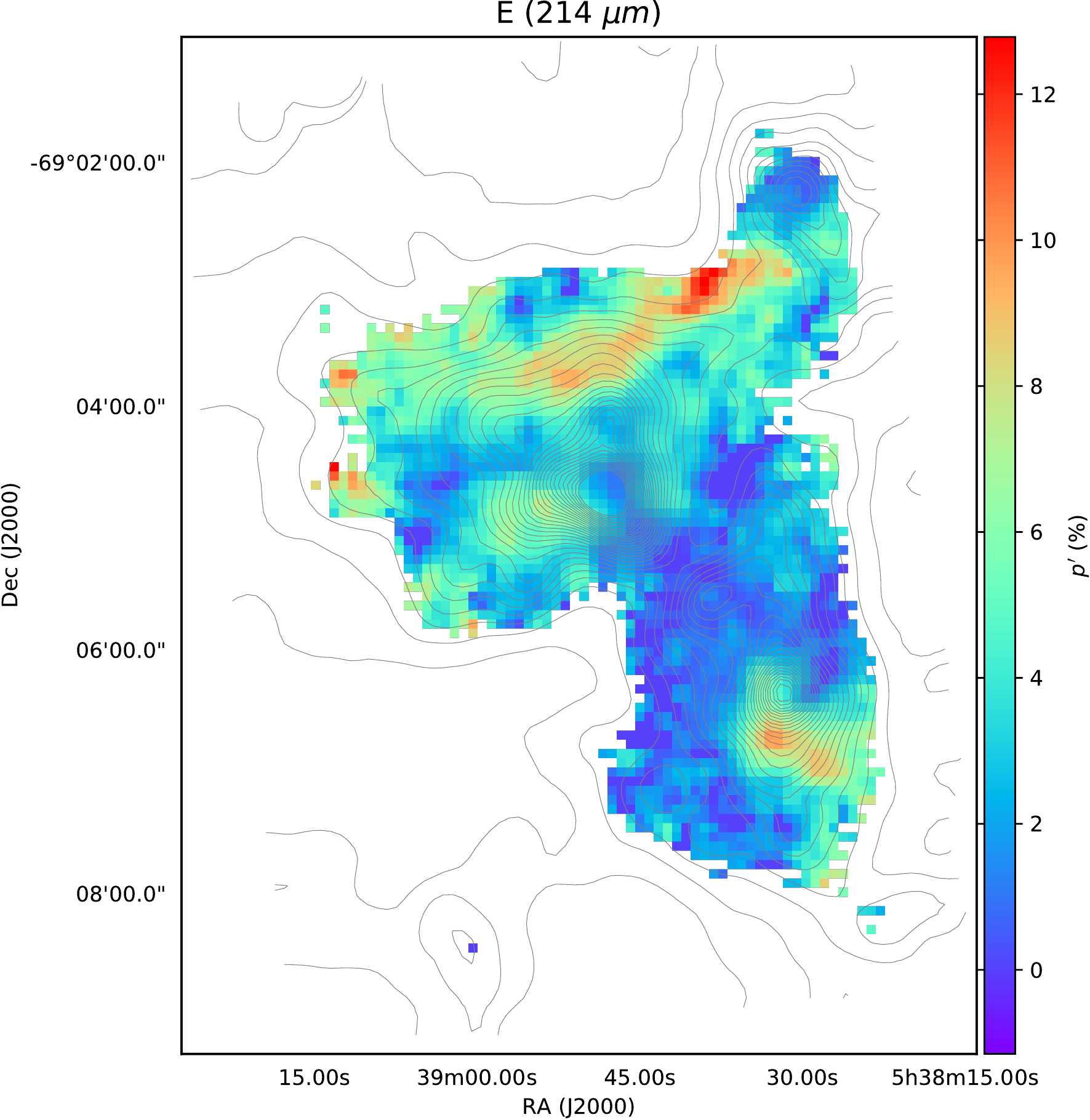}
  \caption{Percentage of polarization (color scale) at 214~\micron\ with overlaid total intensity contours.
    \label{fig:epmap}}
\end{figure}

Another method of visualizing the polarized structure of 30~Dor is shown in Figures~\ref{fig:apmap}--\ref{fig:epmap}. These figures represent the debiased polarization percent $p^\prime$ with overlaid contours of the total intensity, Stokes~$I$.

\begin{figure}[h!]
  \includegraphics[scale = 0.50]{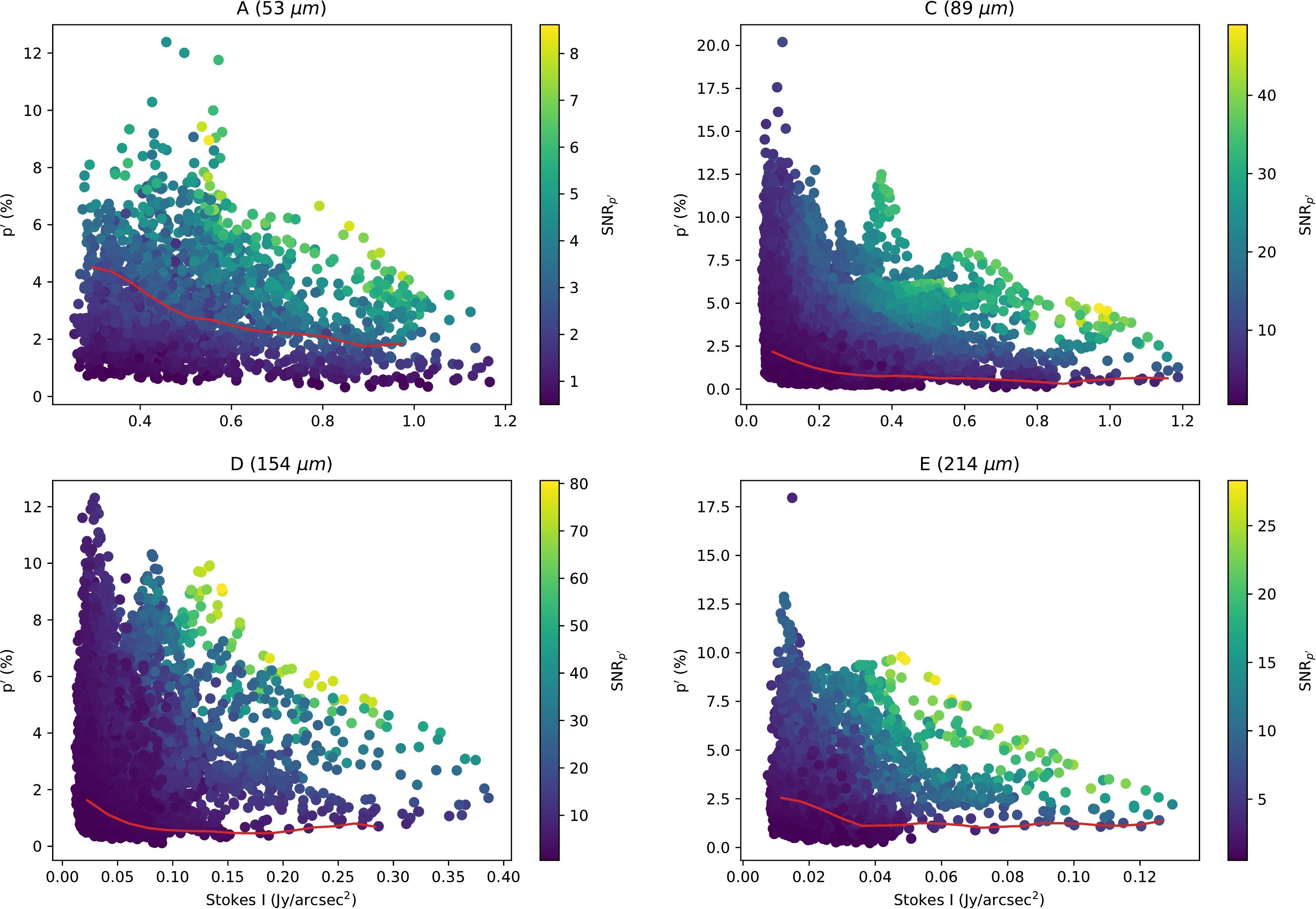}
  \caption{Polarization as a function of  surface brightness at 53~\micron\ (top left), 89~\micron\ (top right), 154~\micron\ (bottom left), and 214~\micron\ (bottom right). All of the data points shown correspond to the SNR$_I > 100$ quality cut in the polarization maps.  The red line shows the approximate envelope for a signal-to-noise cut on $p^\prime$ greater than three---that is, all data points above the red line have a SNR$_{p^\prime}\gtrsim3$.
    \label{fig:PvsI}}
\end{figure}

For those polarization vectors passing quality cuts shown in the above figures, we illustrate the polarization as a function of the surface brightness for each band in Figure~\ref{fig:PvsI}. These plots show a structure more complex than the typical decrease of polarization with increasing surface brightness. We present here these figures to summarize the quality of the data products, but we leave the scientific interpretation for the astronomical community.

\clearpage

\section{Final Remarks}\label{sec:final}

The S-DDT is dedicated to providing the community with high-level, non-proprietary data products for high impact science covering a variety of topics. We have presented here the first program of this initiative with observations of 30~Doradus using the FIR polarimeter, HAWC+. We performed imaging polarimetric observations of the continuum dust emission in four bands in the range of $50-250~\micron$ covering a region of $\sim$6~arcmin$^2$. SOFIA Science Center has delivered science-ready, `Level~4' data, and we have developed and presented tools that showcase both the quality of the observations as well as possible use cases of the data and visualization schemes for the science community. If the users have any questions about these data or future programs, they should contact the SOFIA Helpdesk. In the future, S-DDT program 76\_0003 with polarimetric observations using HAWC+ of galaxies will be released with complementary tools for faint objects.

\acknowledgments

Based on observations made with the NASA/DLR Stratospheric Observatory for Infrared Astronomy (SOFIA) under the SDDT Program. SOFIA is jointly operated by the Universities Space Research Association, Inc. (USRA), under NASA contract NAS2-97001, and the Deutsches SOFIA Institut (DSI) under DLR contract 50 OK 0901 to the University of Stuttgart. SOFIA Program ID: \#76\_0001. 

%

\vspace{5mm}
\facilities{SOFIA (HAWC+)}


\software{astropy \citep{astropy}, APLpy \citep{aplpy}}

\end{document}